# Resistive anisotropy in the charge density wave phase of Kagome superconductor CsV$_3$Sb$_5$ thin films

Han-Xin Lou,[1] Xing-Guo Ye,[1] Xin Liao,[1] Tong-Yang Zhao,[1] An-Qi Wang,[1] Da-Peng Yu,[2,3,4] and Zhi-Min Liao[1,2,a)]

**AFFILIATIONS**

[1]State Key Laboratory for Mesoscopic Physics and Frontiers Science Center for Nano-optoelectronics, School of Physics, Peking University, Beijing 100871, China
[2]Hefei National Laboratory, Hefei 230088, China
[3]Shenzhen Institute for Quantum Science and Engineering, Southern University of Science and Technology, Shenzhen, Guangdong, China
[4]International Quantum Academy, Shenzhen, Guangdong, China

a)Author to whom correspondence should be addressed: liaozm@pku.edu.cn

**ABSTRACT**

We investigate the resistive anisotropy in CsV$_3$Sb$_5$ thin films within the charge density wave phase. Using a device structure with twelve electrodes symmetrically distributed in a circular shape, we measure the resistivity anisotropy by varying the current direction. A twofold resistivity anisotropy modulated by temperature is found, which is fully consistent with the electronic nematicity in CsV$_3$Sb$_5$, that is, the spontaneous rotational symmetry breaking by electronic degree of freedom. Additionally, the resistivity anisotropy also shows modest changes by applying magnetic fields, implying the possible chiral charge orders with time-reversal symmetry breaking. These findings provide deep insights into the correlated electronic states in Kagome materials and highlight the unique properties of CsV$_3$Sb$_5$ in the two-dimensional regime.

The interplay between symmetry breaking and electron interactions is known to spawn a plethora of exotic quantum states and unconventional phases, such as superconductivity,[1,2] ferromagnetic order,[3,4] various density waves, including charge, spin, and pair density waves, as well as Mott insulator.[5–7] Among these, electronic nematicity, i.e., an electronic ordered state with spontaneous breaking of rotational symmetry by electronic degrees of freedom, has attracted keen attention, which is a ubiquitous phenomenon in correlated electron systems such as high-temperature superconductors.[8–11] Of particular interest is that electronic nematicity has also been observed in quasi-two-dimensional Kagome superconductors of type AV$_3$Sb$_5$ (A = K, Rb, and Cs), which exhibit a rich electronic phase diagram, including topological properties, charge density waves (CDWs), and superconductivity.[12–16] These materials are Z$_2$ topological Dirac semimetals that feature Dirac points and van Hove singularities near the Fermi level.[17,18] They exhibit a cascade of phase transitions, leading to exotic electronic orders with symmetry breaking.[16,19–25] Below the CDW transition temperature (T$_{CDW}$ ≈ 70–100 K), a 2 × 2 CDW phase emerges, which triggers electronic nematicity, reducing the rotational symmetry from C$_6$ to C$_2$.[13,19,26–29] At lower temperatures, the system exhibits unconventional superconductivity, including chiral and nodal superconductivity, as well as triplet pairing topological superconductivity.[23,25,30–35] It is found that the intertwining between superconductivity and CDW order can be detected and tuned by electrical and chemical doping.[36–41]

The resistive anisotropy in AV$_3$Sb$_5$ is a sensitive probe of the underlying electronic order and symmetry breaking.[24,26,42–44] It has been reported that in-plane magnetoresistance anisotropy indicates a three-dimensional hidden phase in CsV$_3$Sb$_5$.[43] Previous studies on bulk samples have provided important insights with various experimental techniques,[24,26,43,44] but the behavior of thin films, especially in the two-dimensional regime, remains less explored. By employing a sophisticated device structure with twelve symmetric electrodes, we adopted comprehensive transport measurements to explore the resistivity anisotropy in CsV$_3$Sb$_5$ thin films, less than 30 nm, as a function of current direction, temperature, and magnetic field. The resistive anisotropy measured by rotating the current direction reflects the inherent symmetries of CsV$_3$Sb$_5$ with twofold symmetry breaking directly. This behavior aligns with the emergence of electronic nematicity driven by the CDW state, which has been widely discussed using STM, NMR, Raman, Kerr, and ARPES,[19,27,28,45,46] but rarely by transport measurements. Furthermore, we observe that the resistivity anisotropy is modulated through the application of magnetic fields, which suppress the superconducting state, thus highlighting the competitive interplay between electronic states and supporting the potential chiral charge orders with time-reversal symmetry breaking.

Figure 1 illustrates the schematic diagram of device structure and measurement setup. As shown in Fig. 1(a), the Ti/Au electrodes (2/10 nm thick) were prefabricated on the SiO$_2$/Si substrate. The thin layer CsV$_3$Sb$_5$ flake was obtained from the bulk crystal through mechanical exfoliation and then transferred onto the Ti/Au electrodes. The device is encapsulated by the flake of hexagonal boron nitride (hBN) to prevent oxidation of CsV$_3$Sb$_5$. The measurement configurations are shown in Fig. 1(b), where V$_{xx}$ is the voltage parallel to the current direction and V$_{xy}$ is the voltage perpendicular to the current direction. To investigate the symmetry breaking phenomena in CsV$_3$Sb$_5$, devices with 12 symmetrical Ti/Au electrodes, arranged at 30° of equal step in the shape of a circle, were specifically crafted to quantify the resistivity anisotropy. In the experiment, the current is applied along two opposite electrodes, and the voltage is measured through another pair of electrodes parallel or perpendicular to the direction of the current. The resistivity is obtained through I–V curves. To measure the resistivity under different current directions, the whole measurement framework is rotated together to keep the relative position unchanged [Fig. 1(b)]. In this way, the resistivity anisotropy with current along different directions is studied.

The optical image of a typical CsV$_3$Sb$_5$ device is depicted in Fig. 2(a), the thickness of CsV$_3$Sb$_5$ thin flakes is less than 30 nm. The angle $\theta$ is established as the angular difference between the direction of the electrical current and the electrode pair's baseline. The results of two devices, named device S1 and S2, are presented in this work. All data in the main text come from device S1, and the related data for device S2 are provided in the supplementary material (see Fig. S1 in the supplementary material). An initial measurement of resistivity in relation to temperature was conducted, revealing a distinct superconducting transition at approximately 4 K, marked by a sharp decline in resistivity [Fig. 2(b)]. The superconductivity observed in Kagome lattice materials has been a topic of extensive research and has recently garnered significant attention.[18,23,31,33,47–51] Notably, the gradient of the resistivity–temperature ($\rho$–T) curve experiences an abrupt change, peaking at a maximum d$\rho$/dT at around 75 K,[52,53] as indicated in the inset of Fig. 2(b), aligning with the temperature for the CDW phase transition, T$_{CDW}$ (see Fig. S2 in the supplementary material). The impact of an out-of-plane magnetic field on the resistivity characteristics at 1.8 K is portrayed in Fig. 2(c), with the critical magnetic field, B$_c$, estimated to be ∼1 T. Subsequently, the temperature dependence of resistivity was assessed under varying magnetic fields ranging from 0 to 0.5 T, demonstrating the detrimental influence of magnetic field on superconductivity [Fig. 2(d)].

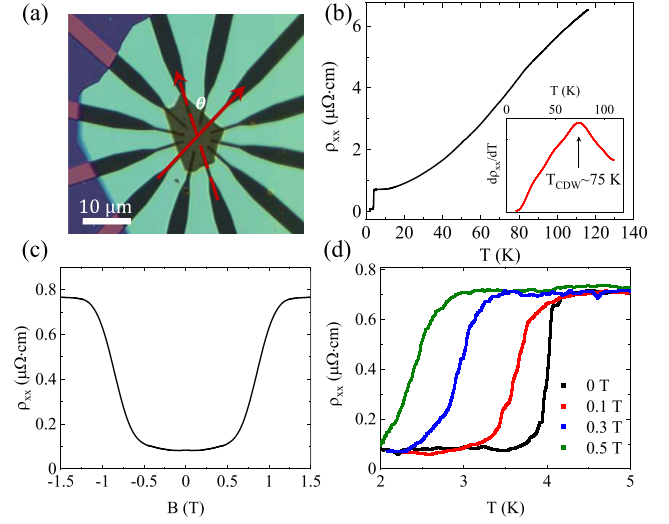

**FIG. 2.** Superconductivity of CsV$_3$Sb$_5$. (a) The optical image of a typical CsV$_3$Sb$_5$ device, and an angle $\theta$ is defined as the angle between the applied current direction and a baseline of the electrode pair (dotted line). (b) Temperature-dependent resistivity measured by four-probe method with T$_c$ ∼ 4 K. The peak of the d$\rho$/dT curve in the inset denotes the onset of charge density wave order at ∼75 K. (c) Magnetic field dependence of resistivity at 1.8 K. (d) Resistivity measured near the superconducting transition temperature under different out-of-plane magnetic fields.

Generally, a series of operations within device fabrication, including crystal growth, mechanically exfoliation of CsV$_3$Sb$_5$ and dry transfer, will inevitably introduce strain in the experimental reality when there is no low-stress designs. While such weak strains are often negligible in conventional compounds, Ref. 44 highlights that pristine CsV$_3$Sb$_5$ is highly unstable and prone to nematic state formation under minimal perturbations. Guo et al. report that strain-free CsV$_3$Sb$_5$ remains isotropic across all temperatures but becomes extremely sensitive to nematicity in the CDW state when exposed to weak disturbances such as strain or magnetic fields.[44] Thus, in our work, without specifically eliminating strain, electronic nematicity and anisotropy may be expected.

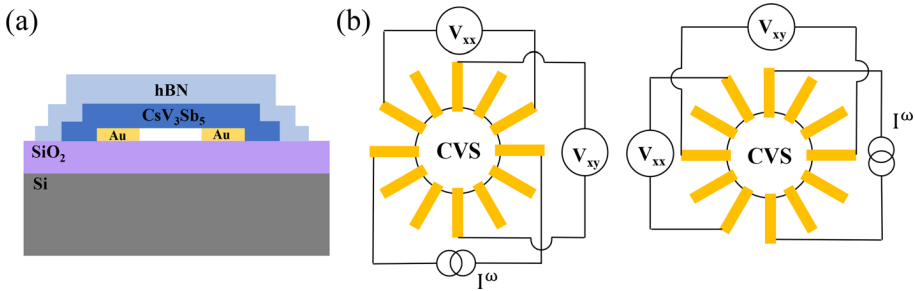

**FIG. 1.** Schematic diagram of device structure and measurement setup. (a) Sideview schematic of CsV$_3$Sb$_5$ structure. CsV$_3$Sb$_5$ is placed on the prefabricated bottom electrodes, encapsulated by the flake of hBN for protection. (b) Schematic of measurement circuit with four-probe measurement method, where the whole measurement framework is rotated while the relative position remains unchanged.

Twelve symmetrical electrodes are used to apply current in various orientations, allowing the measurement of resistivity anisotropy related to nematicity. The current orientation is specified by the angle $\theta$, as shown in Fig. 2(b). Figure 3(a) presents the angle-dependent I–V curves at 5 K, revealing significant resistivity variation with angle and clear twofold symmetry in polar coordinates [Fig. 3(d)]. Notably, for an anisotropic material with spontaneous symmetry breaking, a transverse voltage can be detected when the current is not aligned with the principal axis, as observed in copper oxide superconductors using sophisticated circular electrode configurations.[54] To elucidate the source of resistive anisotropy, the angular dependence of $\rho_{xx}$ and $\rho_{xy}$ is compared in Fig. 3(b) (see also Fig. S3 in the supplementary material). Consistent with previous results,[54] $\rho_{xx}$ and $\rho_{xy}$ display similar oscillations with the same angular period but a phase shift of approximately $\pi/4$. These results indicate that the resistive anisotropy in $CsV_3Sb_5$ is intrinsic, ruling out the effects of contact misalignment or sample heterogeneity.

Based on the twofold symmetry of the system, the angle-dependent resistivity curve of $CsV_3Sb_5$ can be fitted by the following formula:[55]

$$\rho_{xx} = \rho_{2\theta}\sin^2(\theta - \theta_0) + r\rho_{2\theta}\cos^2(\theta - \theta_0), \quad (1)$$

where $\rho_{2\theta}$ is the resistivity of $CsV_3Sb_5$ along the crystalline axis, $\theta_0$ is the misalignment angle between electrodes and the crystalline axis of $CsV_3Sb_5$, and $r$ is the anisotropic ratio. However, the fitting model becomes progressively less accurate at lower temperatures. For example, at 30 K, there is a significant difference ($\Delta\rho$) between the measured resistivity and the fit of Eq. (1), as shown in Fig. 3(c). It is found $\Delta\rho$ shows a clear fourfold symmetry. Therefore, a modification to the formula is required at lower temperatures by incorporating a corrective term that embodies this fourfold symmetry. The revised equation can then be expressed as follows:

$$\rho_{xx} = \rho_{2\theta}\sin^2(\theta - \theta_0) + r\rho_{2\theta}\cos^2(\theta - \theta_0) + \rho_{4\theta}\cos(4(\theta - \theta_0)), \quad (2)$$

where $\rho_{4\theta}$ is the correction term for the fourfold symmetry resistivity. As shown in Fig. 3(d), the measured data at low temperatures now can be well-fitted by this modified model.

The fitting results of anisotropic ratio $r$ and $\rho_{4\theta}$ are presented in Figs. 3(e) and 3(f), respectively. The remarkable feature is that ratio $r$ experiences sharp variations as the temperature decreases below $T_{CDW}$, coincident with the appearance of $\rho_{4\theta}$. It is noteworthy that the nematicity driven by CDW in $CsV_3Sb_5$ is known to cause a $C_2$ anisotropy, which contributes to twofold symmetry anisotropic resistivity. The decrease in $r$ and the appearance of $\rho_{4\theta}$ indicate that the origin of symmetry breaking in the CDW state is different from the lattice distortion-induced twofold symmetry, where $\rho_{4\theta}$ emerges as a result of competition between lattice distortion-induced $C_2$ and charge density wave-induced $C_2$ anisotropy. The presence of the fourfold symmetry term $\rho_{4\theta}$ strengthens the concept of resistivity anisotropy modulated by CDW, combined with the $C_2$ term, serving as a hallmark of the nematic CDW state.[26,43]

In addition to the symmetry breaking induced by intrinsic effect, i.e., the electronical nematicity driven by CDW, there also may exist some extrinsic effects that can induce the anisotropy of resistivity. External effect of contact was carefully discussed and ruled out in the supplementary material (see Fig. S4 in the supplementary material). Remarkably, there are some differences in regard to resistivity anisotropy between our work and Ref. 44, which is due to the different sample thicknesses used. Compared with the bulk samples with thickness above 1 μm in Ref. 44, which is in the three-dimensional regime, the exploited $CsV_3Sb_5$ thin films with thickness less than 30 nm in our work is in the crossover regime between three dimensions and two dimensions. The reduced dimension may lead to enhancement of resistivity anisotropy, since low-dimensional materials are more sensitive to substrate-induced strain or structural phase transition.

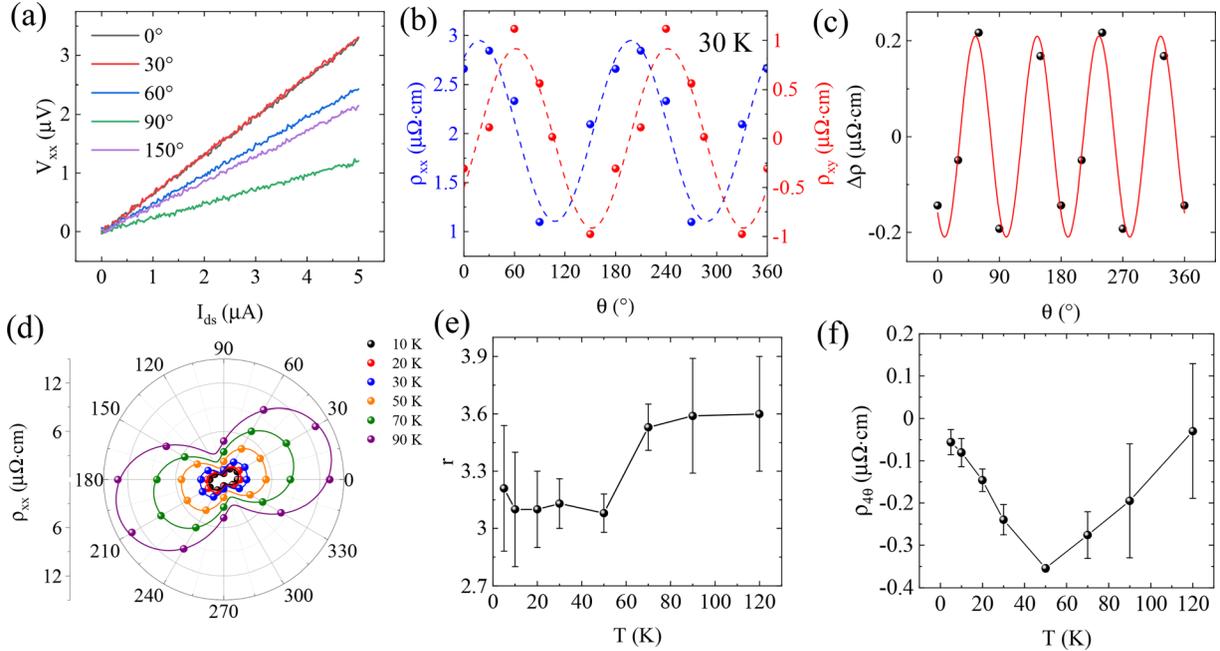

**FIG. 3.** Twofold symmetry of angular-dependent resistivity at different temperatures. (a) I–V curves measured at 5 K on applying current along different angles. (b) Angular-dependent resistivity of $\rho_{xx}$ (blue dots) and $\rho_{xy}$ (red dots) measured at 30 K. (c) $\Delta R$ obtained after subtraction of the fitted curve from Eq. (1) at 30 K. (d) The resistivity curves in polar coordinates, which are well-fitted with an additional term $\rho_{4\theta}$ that introduces fourfold symmetry. Temperature dependence of the (e) resistivity anisotropy ratio $r$ and (f) $\rho_{4\theta}$, both plotted with error bars and derived from the fitting outcomes of Eq. (2).

The occurrence of nematic electronic state[9,10] is acknowledged to be widespread within correlated electron systems, such as those found in high-temperature superconductors[56] and ultraclean quantum Hall systems.[57] In these systems, the crystal structure's rotational symmetry is disrupted. Given its potential to result in superconductivity via an unconventional pairing mechanism, the interconnection between electron nematicity and superconductivity has consistently been an intriguing concept.[58–61] As shown in Fig. 3(f), the amplitude of $\rho_{4\theta}$ decreases with decrease in temperature below 50 K, since the resistivity itself decreases with temperature. As the temperature decreases further, superconducting phases will eventually appear.

Continuous phase transitions are often accompanied by complex interweaving. To delve deeper into the interwoven electronic phases, a magnetic field is introduced while the material is in the superconducting state. Figures 4(a) and 4(e) depict the $\rho_{xx}$ vs B curves under out-of-plane and in-plane magnetic fields at 1.8 K, respectively. It is observed that the critical magnetic field $B_c$ is ∼1 and 6 T for the out-of-plane and in-plane orientations, respectively. Once a magnetic field exceeding the $B_c$ is applied, the anisotropy of resistivity under various magnetic fields is measured. The resulting gourd-shaped curves, when fitted, confirm the presence of twofold symmetry resistivity anisotropy within the $CsV_3Sb_5$, as shown in Figs. 4(b) and 4(f). Most notably, the absolute value of $\rho_{4\theta}$ is observed to increase with increasing magnetic field, as illustrated in Figs. 4(d) and 4(h), while the resistivity anisotropy ratio $r$ shows abrupt changes near $B_c$ [Figs. 4(c) and 4(g)]. Similar experimental results have also been observed in device S2 (see Fig. S5 in the supplementary material). The modulations of resistivity anisotropy, accompanied by the resurgence of the fourfold symmetry term $\rho_{4\theta}$, upon the application of a magnetic field suggest that twofold anisotropic charge order state is intensified, concurrent with the suppression of superconductivity. At low temperatures, the superconducting state is dominant, while the abrupt changes of resistivity anisotropy suggest that there is a competitive relationship between electronic states, which can be modulated by temperature and magnetic field. Moreover, since the resistivity anisotropy is modulated by magnetic field, it may imply the electronic chirality of the nematic CDW state with time-reversal symmetry breaking, which has been theoretically proposed as chiral flux phase or charge bond order with orbital loop current.[41,55,62]

Our results align with the idea that the electron fluid spontaneously breaks crystal symmetry below the charge density wave transition. The resistivity anisotropy is attributed to electronic nematicity for two reasons. First, the anisotropy varies non-monotonically with temperature and is tunable by an out-of-plane magnetic field, which suggests a connection to electronic nematicity rather than lattice distortion. Second, the emergence of a $C_4$ term, $\rho_{4\theta}$, indicates competition between $C_2$ anisotropies from different sources: nematicity and lattice distortion.

In conclusion, we report the temperature and magnetic field-modulated resistivity anisotropy in $CsV_3Sb_5$ and propose a sophisticated device structure: a probe array formed by 12 electrodes arranged around the sample at equal intervals. This arrangement allows currents to be applied along different directions of the sample to detect the resistivity anisotropy of $CsV_3Sb_5$. Therefore, the inherent symmetries of $CsV_3Sb_5$ can be directly reflected by measuring the resistivity anisotropy while rotating the current without applying an external magnetic field. In the experiment, we observed temperature-modulated changes in resistivity anisotropy, exhibiting the twofold symmetry characteristic of nematic CDW in the thin films of $CsV_3Sb_5$. When entering the CDW state by decreasing temperature, the CDW-driven $C_2$ anisotropy competes with the lattice distortion-induced one, leading to the emergence of a fourfold signal. Our work reveals how these correlated electronic states affect the transport properties of $CsV_3Sb_5$ by transport measurements. The CDW-induced resistivity anisotropy diminishes at low temperatures as the material approaches the transition to superconductivity, implying a competitive interaction between the CDW and superconductivity. Upon applying a magnetic field to curtail superconductivity, there is a notable change in resistivity anisotropy, signaling the occurrence of chiral charge orders with time-reversal symmetry breaking. Our findings provide profound insights into the mechanisms driving exotic quantum states in Kagome-based materials.

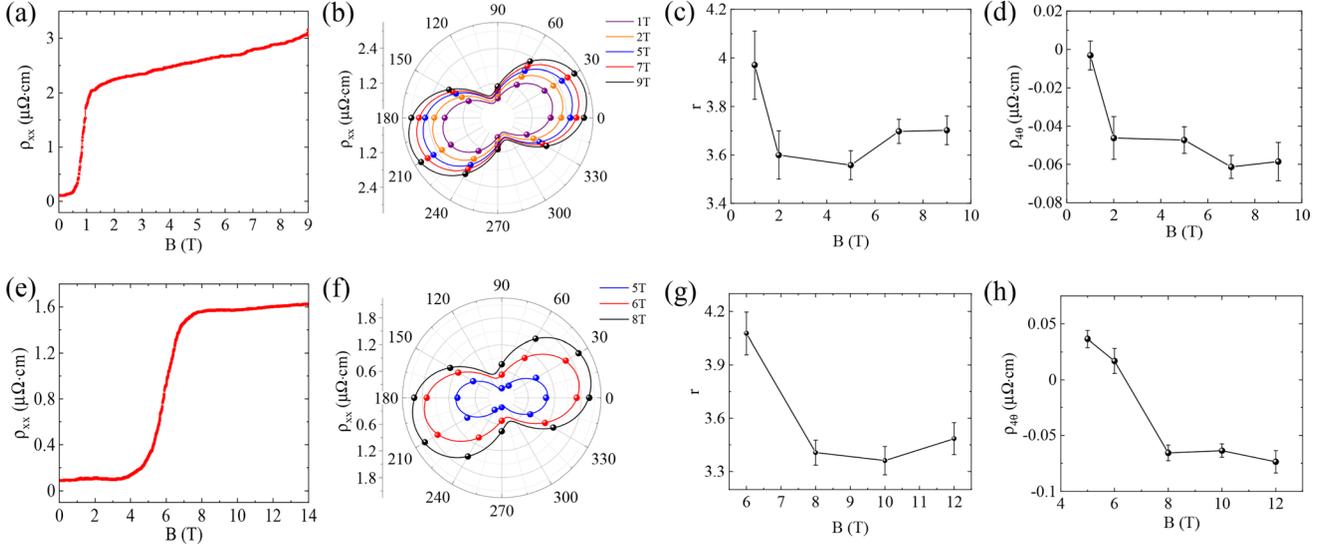

**FIG. 4.** Modulations of resistive anisotropy under magnetic field. The results at 1.8 K on varying (a)–(d) out-plane and (e)–(h) in-plane magnetic field. Panels (a) and (e) show that the critical magnetic fields are 1 and 6 T, respectively. Panels (b) and (f) show that the angle-dependent resistivity curves regain anisotropy following disruption of the superconducting state by applying magnetic fields, with the gourd-shaped curves fitting well to Eq. (2). Panels (c) and (g) show the magnetic field dependence of the resistivity anisotropy ratio . Panels (d) and (h) show the magnetic field dependence of the fourfold symmetry term $\rho_{4\theta}$.

See the supplementary material for more details on device fabrication and on transport measurement method, for additional measurement data for different devices, and for a discussion of the extrinsic effect of contact.

This work was supported by the Innovation Program for Quantum Science and Technology (Grant No. 2021ZD0302403) and the National Natural Science Foundation of China (Grant Nos. 61825401 and 62321004).

# Supplementary Materials for:

# Resistive Anisotropy in the Charge Density Wave Phase of Kagome Superconductor $CsV_3Sb_5$ Thin Films

Han-Xin Lou[1], Xing-Guo Ye[1], Xin Liao[1], Tong-Yang Zhao[1], An-Qi Wang[1], Da-Peng Yu[2,3,4], Zhi-Min Liao[1,2*]

[1] State Key Laboratory for Mesoscopic Physics and Frontiers Science Center for Nano-optoelectronics, School of Physics, Peking University, Beijing 100871, China.
[2] Hefei National Laboratory, Hefei 230088, China.
[3] Shenzhen Institute for Quantum Science and Engineering, Southern University of Science and Technology, Shenzhen, Guangdong, China.
[4] International Quantum Academy, Shenzhen, Guangdong, China.
*E-mail: liaozm@pku.edu.cn

**Methods**

**Device Fabrication.** Employing the standard mechanically exfoliated method, we obtained $CsV_3Sb_5$ thin films, with thickness less than 30 nm, from the bulk crystals. The Ti/Au electrodes (2/10 nm thick) were prefabricated on the $SiO_2$/Si substrate by electron beam evaporation, and the $CsV_3Sb_5$ thin films were transferred onto the electrodes using a polymer-based dry transfer technique. The device is encapsulated by the flake of hexagonal boron nitride (hBN) to prevent oxidation of $CsV_3Sb_5$. The whole exfoliating and transfer process was done in an argon-filled glove box with $O_2$ and $H_2O$ content below 0.01 parts per million to avoid sample degeneration.

**Transport Measurements.** Transport measurements were carried out in an Oxford cryostat with a variable temperature insert and a superconducting magnet. First-harmonic signals were collected by standard lock-in techniques (Stanford Research

Systems Model SR830) with frequency ω. Frequency $\omega$ equals 17.777 Hz unless otherwise stated. The direct current was applied through Keithley 2400 SourceMeters.

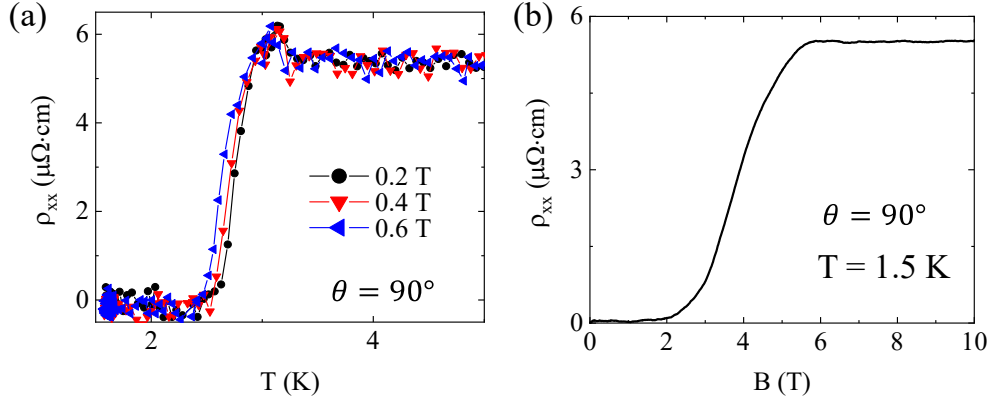

**FIG. S1. Superconductivity of device S2.** (a) The temperature-dependent resistivity is measured with current along $\theta = 90°$ under different out-of-plane magnetic fields, showing resistivity terribly close to zero below the superconducting transition temperature, even with current along $\theta = 90°$ (the maximum resistivity direction). (b) Magnetic field dependence of resistivity at 1.5 K with current along $\theta = 90°$, demonstrating the destructive effect of the magnetic field on superconducting state.

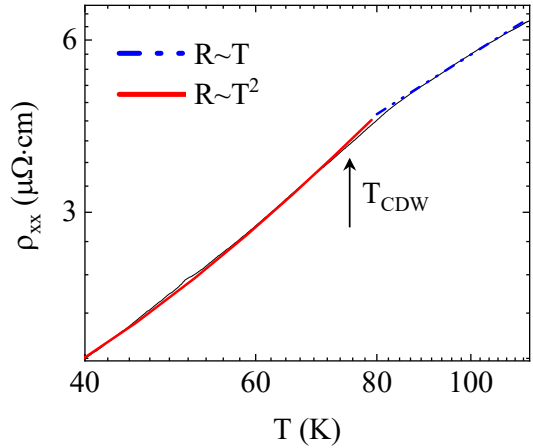

**FIG. S2. Temperature-dependent resistivity measured by four-probe method in device S1.** At high temperature, the resistivity of $CsV_3Sb_5$ is proportional to the temperature, in which the R~T curve (blue dotted line) is well fitted. As the temperature

drops, a transition to R~T² (red solid line) is observed in the CsV$_3$Sb$_5$, with the turning point at 75 K. The temperature-dependent transition of the resistivity indicates the CDW phase transition, with the CDW transition temperature of 75 K.

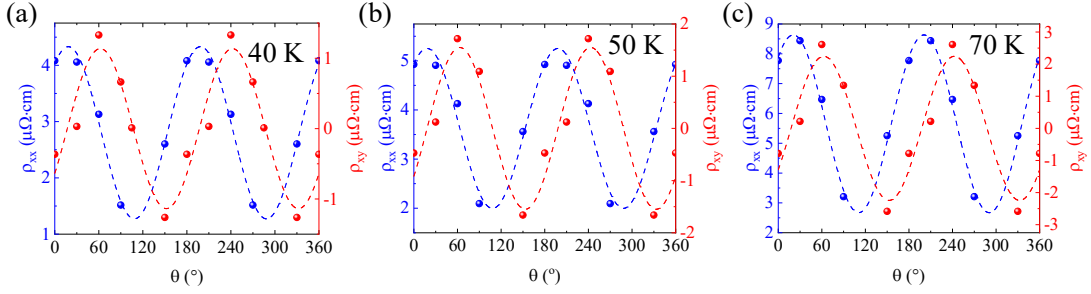

**FIG. S3. Longitudinal and transverse resistivity in device S1 at different temperatures.** The longitudinal resistivity $\rho_{xx}$ is fitted by $\rho_{xx}(\theta) = r\rho_0 cos^2(\theta - \theta_0) + \rho_0 sin^2(\theta - \theta_0)$, and the transverse resistivity $\rho_{xy}$ is fitted by $\rho_{xy}(\theta) = \rho_0(1-r)sin(\theta - \theta_0)cos(\theta - \theta_0)$. Both of $\rho_{xx}$ and $\rho_{xy}$ presents the similar oscillations with same period, but a phase shift around $\pi/4$.

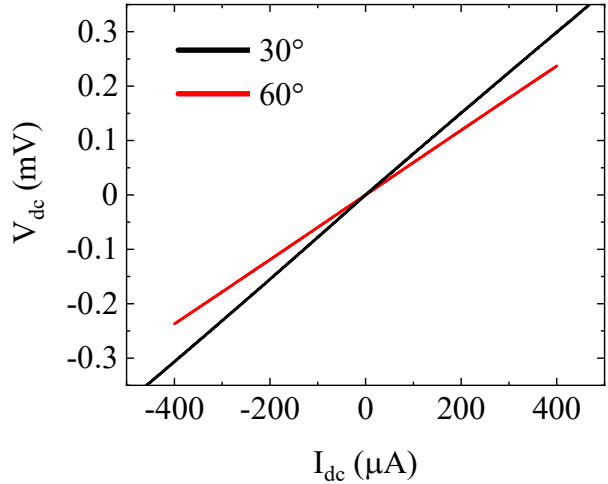

**FIG. S4. Examples of I-V characteristics in device S1 at 5 K.** The R-square of all the linear fittings is at least larger than 0.9999, indicating perfect linearity.

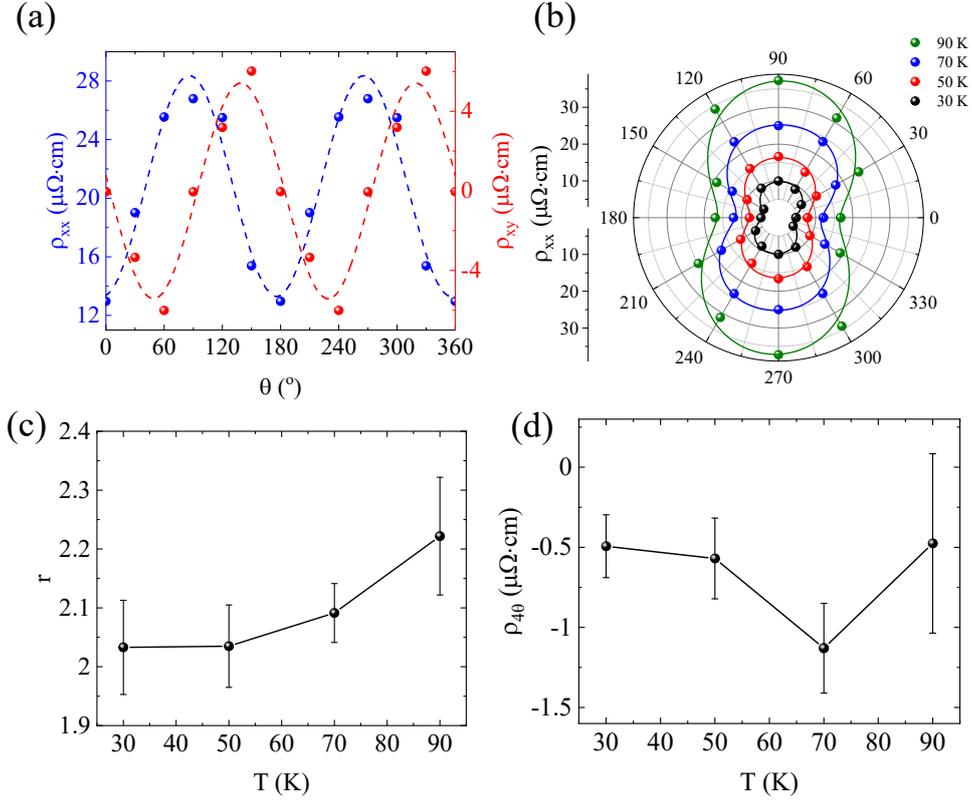

**FIG. S5. Reproducible results in device S2.** (a) Angular-dependent resistivity of $\rho_{xx}$ and $\rho_{xy}$ measured at 70 K in device S2. The longitudinal resistivity $\rho_{xx}$ presents the similar oscillations with same period to transverse resistivity $\rho_{xy}$, but a phase shift around $\pi/4$, consistent with the results of device S1. (b) Angular-dependent resistivity plotted in polar coordinate measured from 30 to 90 K, fitted with Eq. (2), showing the twofold symmetry, consistent with the device S1. (c)-(d) Temperature dependence of the resistivity anisotropy ratio $r$ and $\rho_{4\theta}$.

In addition to the symmetry-breaking induced by intrinsic effect, *i.e.*, the electronic nematicity driven by the charge density wave (CDW), there also may exist some extrinsic effects that can induce the anisotropy of resistivity. Here we carefully exclude the extrinsic effect of contact.

(1) As shown in Fig. 1(b), the four-probe method is adopted in the measurement to suppress the contact resistance. We have performed current-voltage (I-V) characteristics in device S1. If the contact effect exists, nonlinearity due to the contact-induced diode is expected in the I-V curves. However, as shown in Fig. S4, linear behavior is observed in the I-V curves. The R-square of all the linear fittings is at least larger than 0.9999, indicating perfect linearity. Moreover, such linearity is present under current as high as 400 μA. Note the resistivity in our sample is generally measured under current less than 5 μA (see Fig. 3(a)), which rule out the possible contact imperfections and also avoid the device self-heating.

(2) The deviation from linearity in the I-V curves is obtained by subtracting the linear-dependent part. It is found the deviation part $\Delta V$ is less than 0.1 μV when $|I| < 5$ μA. Compared to Fig. 3(a), the resistivity contributed by the nonlinearity is less than 5%. However, the resistivity ratio between maximum resistivity and minimum resistivity is larger than 200%. This observation also rules out the contact effect as the main case of measured resistivity anisotropy.

(3) We have observed the two-fold symmetry resistive anisotropy in more than one sample, the repeatable regularity of which suggests that this is not due to the inhomogeneity of the electrical contact points. Moreover, the shape of two samples is different, so the influence of sample shape on the resistivity anisotropy can be excluded.

(4) The appearance of the 4θ term occurs only in the CDW state and after the

magnetic field breaks the superconducting state. If it is caused by the contact effect, it should not be significantly modulated with the system phase transition (at least not in a non-monotonic way as in Fig. 3(f)). Moreover, the $\rho_{4\theta}$ shows a non-monotonic dependence on temperature, where maximum $\rho_{4\theta}$ is observed at an intermediate temperature ~ 50 K. Such observation is inconsistent with the contact effect.

(5) For anisotropic materials with spontaneous symmetry breaking, a transverse voltage can be measured when the current is not aligned with the principal axis. As shown in Fig. 3(b), $\rho_{xx}$ and $\rho_{xy}$ present the similar oscillations with same period, but a phase shift around $\pi/4$, which demonstrates that the resistivity anisotropy in the $CsV_3Sb_5$ device is inherent, excluding the possibility that the transverse voltage is caused by contact misalignment or random sample inhomogeneities.